\documentclass[9pt,twocolumn,twoside]{pnas-new}

\usepackage{xcolor}

\usepackage{amsmath} 
\usepackage{amsthm} 
\usepackage{amssymb}	
\usepackage{physics}

\DeclareMathOperator*{\argmin}{arg\,min}

\templatetype{pnasresearcharticle} 

\title{Topological characterization of the continuum of allosteric response}

\author[a,b]{Jason W. Rocks}
\author[b]{Eleni Katifori}
\author[b,1]{Andrea J. Liu} 

\affil[a]{Department of Physics, Boston University, Boston, MA 02215, USA}
\affil[b]{Department of Physics and Astronomy, University of Pennsylvania, Philadelphia, PA 19104, USA}

\leadauthor{Rocks} 

\significancestatement{Allostery is a common regulatory mechanism in proteins where a ligand binding to one site controls the activity at a second distant binding site. Allostery is frequently characterized by mechanical signals transmitted by conformational changes that vary dramatically between different proteins. Here, we introduce a quantitative topological description that unifies all allosteric deformations into a single framework. Our analysis identifies and quantifies the significance of both hinge domains and allosteric pathways, key structural features long associated with allostery. To develop and test the analysis, we use ensembles of networks designed with allostery-like functions and then apply the framework to a variety of allosteric proteins. We show that allosteric deformations fall on a continuum and cannot be classified into distinct archetypes.
}

\authorcontributions{Author Contributions: J.W.R., E.K., and A.J.L. designed research and wrote the paper. J.W.R. performed research, contributed new analytic tools, performed numerical calculations, and analyzed data.}
\authordeclaration{The authors declare no conflict of interest.}
\correspondingauthor{\textsuperscript{1}To whom correspondence should be addressed. E-mail: ajliu@upenn.edu}


\begin{abstract}
Allosteric regulation in proteins is often accompanied by conformational changes that facilitate transmission of mechanical signals between distant ligand binding sites. Typically, these deformations are classified in terms of specific archetypes, including various types of hinge mechanisms or allosteric pathways  localized to sequences of amino acids.
However, many allosteric deformations resist such strict categorization. Here, we introduce a quantitative topological description of allosteric deformation, unifying all archetypal mechanisms into a single framework. The topological description aligns with two key structural features often associated with allosteric deformations, namely hinge domains and allosteric pathways, enabling us to quantify the significance of each of these features. To develop the analysis, we tune computer-generated mechanical networks to perform allostery-like functions, obtaining an ensemble of networks that establish a range of possible allosteric deformations. The analysis shows that these networks' allosteric mechanisms cannot be described in terms of discrete archetypes -- they fall on a continuum. We then apply the same analysis to a collection of allosteric proteins with similar results, showing that our framework encompasses these proteins as well as designed allosteric networks. Our results provide a new picture for allostery, demonstrating not only how it can be described quantitatively, but also giving insight into how it emerges as a collective property.

\end{abstract}

\dates{This manuscript was compiled on \today}

\begin{document}

\maketitle
\thispagestyle{firststyle}
\ifthenelse{\boolean{shortarticle}}{\ifthenelse{\boolean{singlecolumn}}{\abscontentformatted}{\abscontent}}{}


\dropcap{A}llosteric Allosteric regulation of a protein occurs when a molecule binding to one site affects the activity at a second distant binding site~\cite{Ribeiro2016, Thirumalai2019, Wodak2019}.
In many proteins, this intra-molecular communication is transmitted mechanically via specific conformational changes~\cite{Daily2007}.
Currently, numerous methods exist to analyze these deformations and identify key structural features responsible for allosteric function.
However, most of these analyses limit their focus to a single class of archetypal motions, with prominent examples of such archetypes including hinge domains~\cite{Shatsky2004, Keating2009, Taylor2014}, shear mechanisms~\cite{Taylor2014, Mitchell2016}, 
and allosteric pathways~\cite{Feher2014, Dokholyan2016}. As a result, it remains unclear whether such archetypes span the full range of conformational deformations involved in allostery, or even whether such deformations can be neatly classified according to a set of discrete archetypes. 

Here, we introduce a unifying framework to quantify the deformations that accompany allostery in proteins. 
Based on persistent homology, this framework characterizes allosteric motion in terms of topologically significant structural features.
Recently, a similar approach was developed to understand similar behavior in flow networks, revealing the key structures responsible for function~\cite{Rocks2020, Rocks2021}.
In this work, we generalize that approach to three-dimensional mechanical networks.
In particular, we focus on central-force spring networks tuned to perform analogous allostery-like functions~\cite{Flechsig2017, Rocks2017, Tlusty2017, Yan2017, Yan2018, Rocks2019}, which are known to exhibit a wide range of conformational deformations including shear, hinge and twist mechanisms~\cite{Yan2018}, along with more exotic mechanical deformations that utilize low energy edge modes to amplify input responses~\cite{Yan2017}.
To systematically explore this zoo of conformational changes, 
we first generate a large statistical ensemble of networks tuned to exhibit similar allosteric responses.
We then analyze the topological structure of the deformation of each network using persistent homology, obtaining a host of features that fall into two topologically-derived families. One family consists of many candidate hinge domains while the other consists of strain pathways connecting every pair of nodes. By design, these two families align with two key structural features that have long been associated with allosteric deformations: hinge domains and allosteric pathways. A key feature of our analysis is that it provides \emph{quantitative measures} of the topological significance, or scale, of each feature in both families. For each system, we focus on the hinge domain and strain pathway of highest significance and show that these two features and their scales characterize the allosteric response.

We find that conformational changes associated with allostery lie on a continuum in terms of these measures of hinge domain and allosteric pathway significance, rather than falling into distinct clusters with each cluster corresponding to a different archetypal mechanism.  
In particular, hinge domains and allosteric pathways can exist simultaneously, with allosteric pathways often lying along the boundaries between hinge domains.
Furthermore, we show that these features can be used to identify nodes in a network that disrupt allosteric function when they are perturbed.
Finally, we go beyond computer-generated mechanical networks to apply our analysis to a collection of allosteric proteins chosen from the Protein Data Bank (PDB)~\cite{Berman2000}. We find that their conformational changes during allosteric response are markedly similar to those of our synthetic networks, demonstrating that proteins can be characterized using the same unifying framework.

\begin{figure}[t!]
\centering
\includegraphics[width=\linewidth]{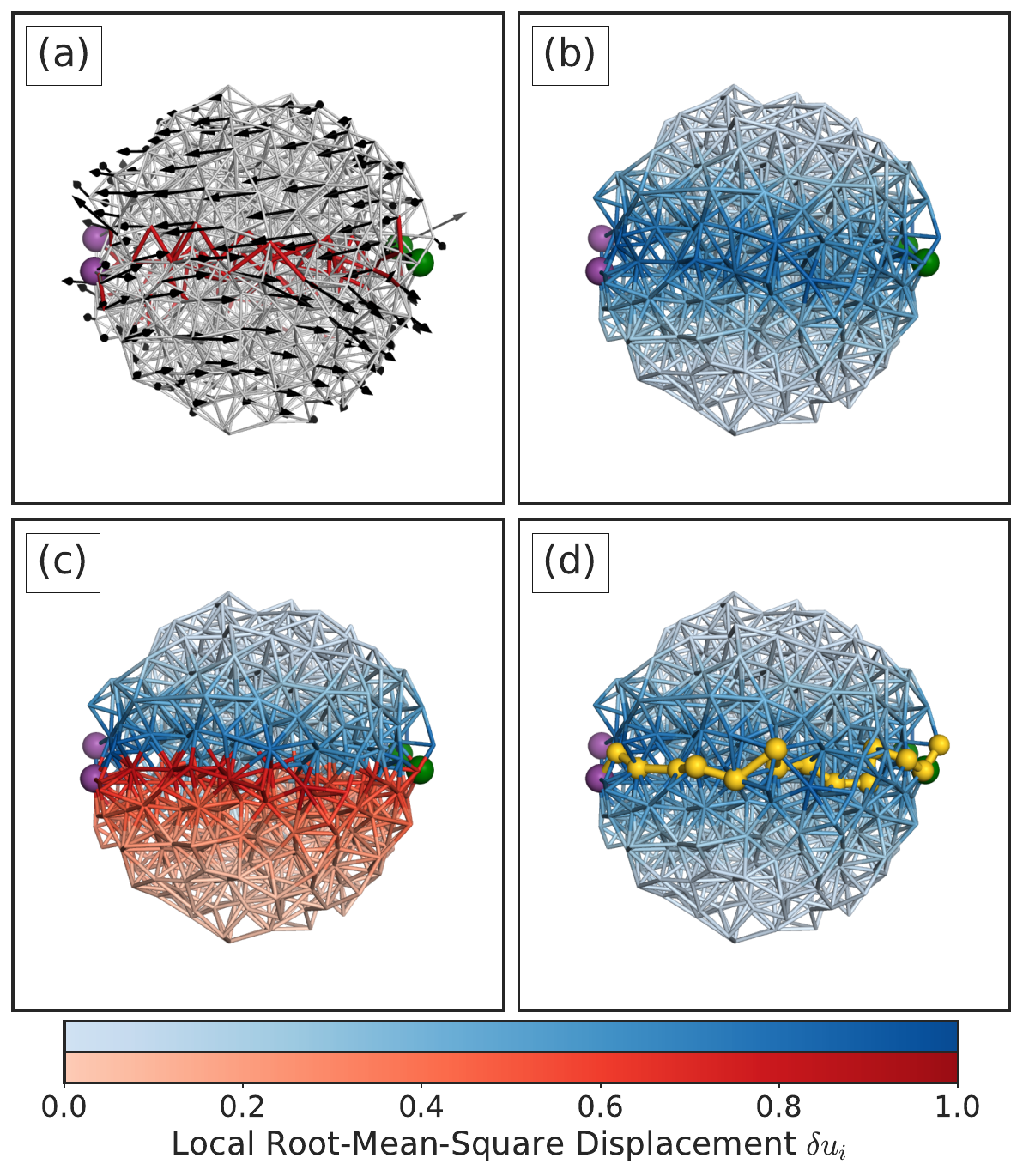}
\caption{
Random spring network tuned to exhibit an allosteric response.
When a unit strain of $\varepsilon_S = 1$ is applied to the two source nodes (shown in purple), the two target nodes (shown in green) respond with a strain of at least $\varepsilon_T \leq -1$.
(a) The structure of the tuned network with removed edges shown in red and the resulting node displacement depicted with black arrows.
(b) The local root-mean-square displacement $\delta u_i$ around each node $i$ is shown in shades of blue.
(c) The local root-mean-square displacement of the two hinge domains identified by our topological analysis are shown in shades of red and blue, color-coded for each domain.
(d) The allosteric pathway is shown in gold, consisting of the nodes with largest $\delta u_i$ that connect the source and target.
}
\label{fig:intro}
\end{figure}

\section*{Tuned Allosteric Networks}

To create an ensemble of mechanical networks that each exhibit a similar allostery-like response, we first generate a collection of three-dimensional disordered central-force spring networks and then tune each one by adjusting its spring stiffnesses~\cite{Rocks2017,Rocks2019}.
Here, we briefly summarize this process and refer the reader to the Methods for complete details.
We derive each initial spring network from the contact network of a randomly generated three-dimensional packing of soft spheres with periodic boundary conditions, created using standard jamming algorithms~\cite{OHern2003, Liu2010, Goodrich2016}. 
We then create roughly spherical networks by removing all nodes (along with their associated edges) outside a specified cutoff radius.
To ensure each network is sufficiently rigid, we also remove any underconstrained nodes on the external surface of our networks and introduce very weak springs between second-nearest neighbors.
In the final ensemble, each network has on average 509 $\pm$ 9 nodes with 2070 $\pm$ 37 nearest neighbor bonds (see SI for distributions).

Next, we tune each network to exhibit an allosteric linear response, so that a specified pair of target nodes exhibits a negative strain of at least $\varepsilon_T \leq -1$ when a positive strain $\varepsilon_S$ is applied across a separate pair of source nodes. In linear response, the target strain $\varepsilon_T$ is simply proportional to the source strain $\varepsilon_S$, so for convenience, we use units where $\varepsilon_S=1$. For each network, we choose the pairs of source and target nodes randomly such that they are located on opposite poles of the network. 
To achieve a desired target strain, we tune the response indirectly by maximizing the target \textit{stress}.
This choice ensures that the overall energy of deforming the target is lowered upon activation of the source, emulating the change in binding energy at the active site of an allosteric protein.
Our tuning scheme utilizes a greedy algorithm: in each step we either remove or add a single edge, modifying the edge that maximally increases the magnitude of the target stress at each step. 
We repeat this greedy edge removal process until the target strain is achieved, excluding networks from our analysis that converge a strain value below the desired threshold.
In Fig.~\ref{fig:intro}(a), we show an example of a spring network tuned in this way.
We find that our strategy of indirectly tuning the target strain by maximizing the target stress yields networks that each exhibit a similar target strain, but with greatly varying amounts of stress communicated from the source to the target.
As we demonstrate, this allows us to achieve tuned networks that exhibit a wide variety of allosteric motions.

\begin{figure*}[t!]
\centering
\includegraphics[width=\linewidth]{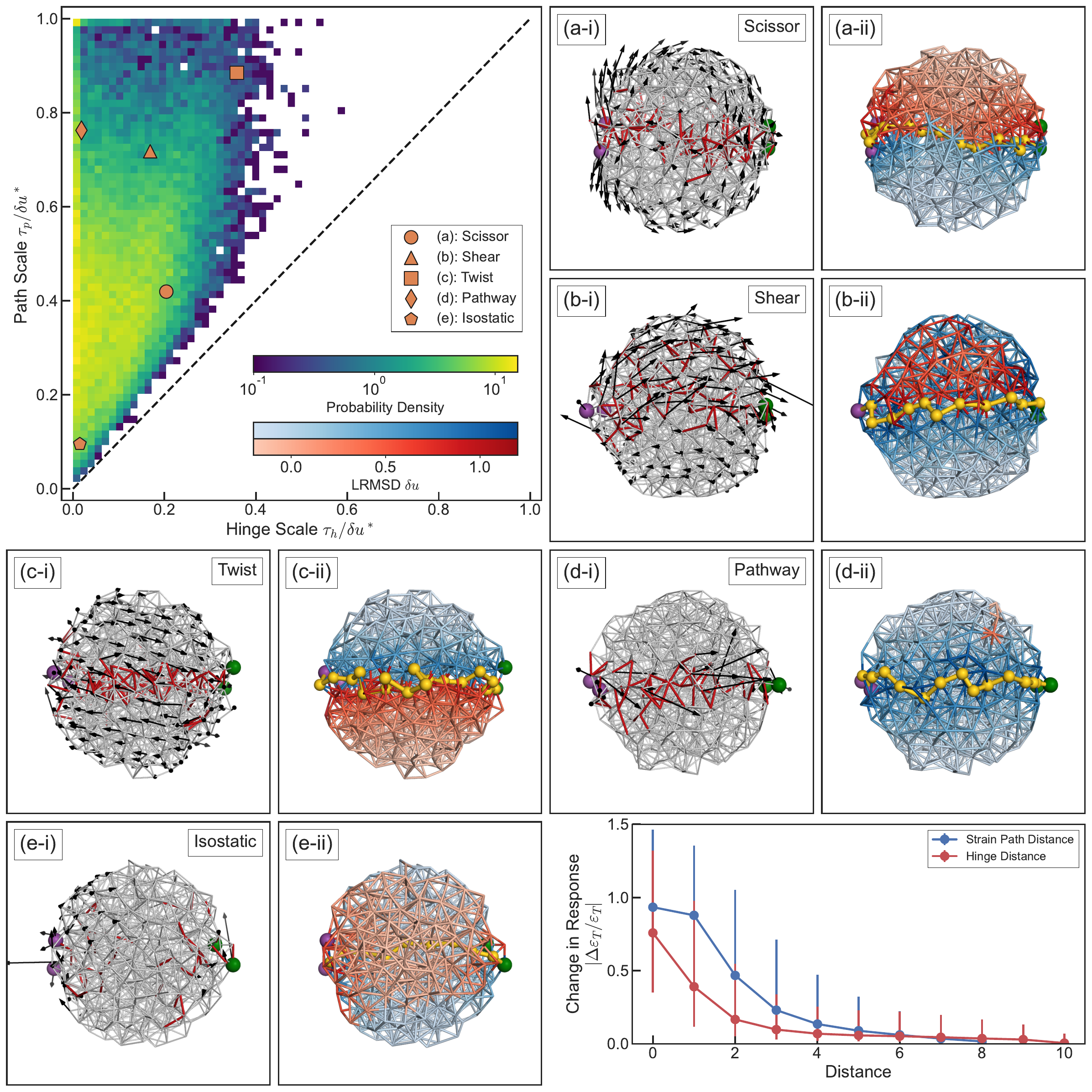}
\caption{The conformational phase space of allosteric motions observed in random spring networks tuned to exhibit allosteric responses. (Top Left) The observed probability density of hinge scales $\tau_h$ and path scales $\tau_p$ exhibited by about 35000 tuned networks. The axes are in units  of the normalization scale $\delta u^*$, taken to be the value of $\delta u_i$ experienced by the source node in the identified allosteric pathway for each network.
A variety of networks with different archetypal motions are observed (indicated by orange points in the distribution) including hinges with (a) scissor, (b) shear, and (c) twist motions, (d) prominent allosteric pathways, and (e) isostatic responses.
For each network, the network structure and displacement are shown on the left, with the local deformation $\delta u_i$, hinge domains, and allosteric pathway shown on the right.
(Bottom Right) The robustness of the tuned response of the target nodes to changes in local structure near the allosteric pathway and hinge boundary.
The magnitude of the fractional change in target response $|\Delta \varepsilon_T / \varepsilon_T|$ is measured when the local neighborhood around a node is constrained to undergo no local deformation.
This quantity is plotted as a function of distance from the allosteric pathway (blue) or the hinge domain boundary (red). Each point represents the median change in response with error bars indicating the interquartile range.
 }
\label{fig:zoo}
\end{figure*}

\section*{Topological Characterization}

We characterize the allosteric motions of our tuned networks in terms of two classes of structural features: 
hinge domains, collections of nodes that move in concert as approximate rigid bodies, 
and allosteric pathways, contiguous chains of adjacent nodes that carry the mechanical response from the source to the target.
While these qualitative descriptions may appear simple, constructing formal quantitative definitions of these features poses a challenge due to the nature of the discrete network response. 
In particular, networks that exhibit hinge-like motions often do not exhibit a clear and distinct boundary between apparent hinge domains.
Similarly, allosteric pathways often do not exhibit uniform deformations between the source and target or may be wider than a single node.
In both cases, we desire a method to identify unique features without the need for choosing arbitrary cutoffs in the local deformations of the response.

To accomplish this goal, we use methods derived from persistent homology, a class of techniques that can discern topological features in topologically and/or geometrically structured data~\cite{Edelsbrunner2010, Otter2017}.
This type of analysis provides a systematic means of identifying features at all scales encoded in a function (the mechanical response) defined in some space (the network nodes and edges).
The approach we use is a generalization of a technique we recently developed to identify topological features in flow networks tuned to deliver a specified pressure drop to a designated target edge in response to a pressure drop at a source edge~\cite{Rocks2020, Rocks2021}.

To apply our analysis, we first measure the local deformation $\delta u_i$ in the neighborhood around each node $i$, using the node displacements of the tuned response to calculate the local root-mean-square displacement (LRMSD) (see Methods). 
In Fig.~\ref{fig:intro}(b), we depict the LRMSD of the tuned network in Fig.~\ref{fig:intro}(a).
Next, we capture the topological structure of the response by performing both an ascending and descending filtration of $\delta u_i$. Because $\delta u_i$ is associated with node $i$, the ascending and descending filtrations yield two distinct spanning trees, which we call the hinge and pathway skeletons, respectively.
The hinge skeleton consists of all edges comprising paths of steepest descent from each node to a local minimum in $\delta u_i$, 
while the path skeleton consists of all paths of steepest ascent to a local maximum (see Methods).

In both cases, these spanning trees provide many candidate features from which to choose.
To identify possible pairs of hinge domains, 
we use the fact that the hinge skeleton naturally decomposes the network into a potentially large number of sectors of relatively rigid motion composed of all nodes whose paths of steepest descent lead to the same local minimum (see SI for example). 
At the same time, each edge in the hinge skeleton that connects a pair of adjacent sectors represents a boundary of relatively large deformations separating them.
Because the hinge skeleton is a tree, removing any one of these boundary edges splits the network into two connected components, effectively combining, or coarse-graining, the sectors to obtain a pair of hinge domains.
We consider each pair of coarse-grained sectors created in this way as a possible pair of hinge domains describing the allosteric motion.
On the other hand, the pathway skeleton contains a unique path connecting each pair of nodes with the property that along such a path, the minimum value of $\delta u_i$ encountered is as large as possible. 
Because the source and target may each consist of multiple nodes, 
each path between a single source node and a single target node within the pathway skeleton represents a possible allosteric pathway (see SI for example).

A key feature of this approach is that it provides a natural measure of topological significance via the scale of each candidate feature (closely related to persistence in persistent homology, see Methods).
To assess the significance of an allosteric pathway in the pathway skeleton, we define the \textit{path scale} $\tau_p$ as the smallest value of $\delta u_i$ encountered along a candidate pathway.
We also define a normalization scale $\delta u^*$ as the value of $\delta u_i$ of the source node at the beginning of the pathway.
The ratio $\tau_p/\delta u^*$ thus measures the fractional drop-off in LRMSD from the source along the pathway.
The larger this ratio, the more significant or pronounced the pathway.
If a continuous pathway of relatively large deformations does not exist between the source and target, then the ratio is small, $\tau_p/\delta u^* \ll 1$,
while a contiguous pathway of deformations comparable to the source will result in a ratio closer to unity,  $\tau_p/\delta u^* \sim 1$.
From among all candidate pathways between the source and target nodes in the pathway skeleton, we choose the pathway with the largest value of $\tau_p$.

Similarly, we assess the significance of a pair of hinge domains by defining the \textit{hinge scale} $\tau_h$.
To calculate this quantity, we measure the minimal $\delta u_i$ along the boundary between the hinge domains, along with the minimum value of $\delta u_i$ observed in each domain. We define $\tau_h$ as the difference between the measured boundary $\delta u_i$ and the maximum of the minima in the two domains (see Methods).
The larger $\tau_h$, the more significant the separation between the hinge domains, 
i.e., the more they tend to appear as regions that each move independently as approximate rigid bodies.
For example, a network lacking a hinge mechanism that exhibits uniform motion will result in a small hinge scale, $\tau_h/\delta u^* \ll 1$,
while a network that exhibits a hinge mechanism consisting of two separate rigidly moving regions will exhibit a larger value, $\tau_h/\delta u^* \sim 1$.
Our analysis always chooses the pair of hinge domains that result in the largest value of $\tau_h$. 

\begin{figure*}[t!]
\centering
\includegraphics[width=\linewidth]{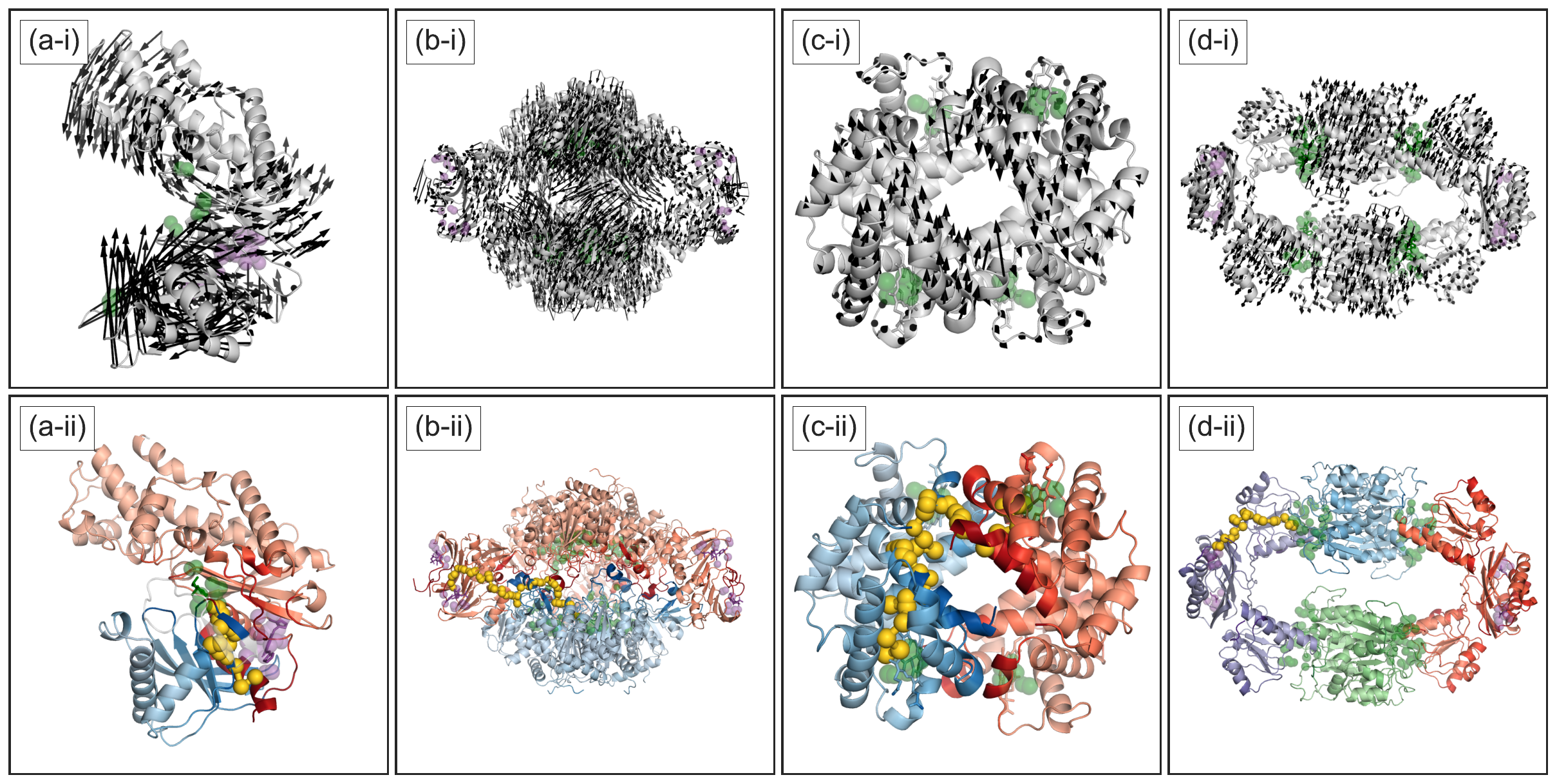}
\caption{Topological analysis of allosteric proteins. (Top Row) Protein configurations in the undeformed state (unbound to substrate) with arrows indicating displacements induced by (an) effector ligand(s) to the active site(s). 
(Bottom Row) Protein configurations in the deformed state (bound to substrate) with color indicating LRMSD. 
Hinge domains are indicated with different colors.
Sequences of atoms creating allosteric pathways are highlighted in gold. 
(a) Glucokinase in its inactive unbound state (PDB ID: 1V4T) and active states (PDB ID: 1V4S) bound to glucose and an artificial activator. (b) ATCase in its active state (PDB ID: 1D09)  bound to a bisubstrate analogue and inactive state (PDB ID: 1RAC) with GTP acting as an inhibitor. (c) Hemoglobin in its inactive state (PDB ID: 2DN2) and active state (PDB ID: 2DN1) bound to four oxygen molecules. A cooperative pathway is shown between two of the active sites. (d) PGDH in its inactive state (PDB ID: 1PSD) bound to serine and in its active state (PDB ID: 1YBA) bound to NAD(+) cofactor. The motion is well described by four hinge domains.}
\label{fig:proteins}
\end{figure*}

\section*{Continuum of Allosteric Motions}

In Fig.~\ref{fig:zoo}, we report the results of applying our topological analysis to an ensemble of about 35000 tuned three-dimensional allosteric networks of roughly similar spherical shape and size.
In the upper left panel, we depict a two-dimensional histogram of the conformational phase space spanned by the networks as a function of the hinge scale $\tau_h$ and path scale $\tau_p$ (normalized by $\delta u^*$).
The color of each pixel represents the number of networks that fall in that bin, indicating the observed probability of each combination of hinge and path scales.
Most strikingly, we observe that rather than falling into discrete clusters, tuned allosteric mechanical networks span a continuum of responses, often exhibiting both hinges and allosteric pathways of varying scales.
In particular, scissor, shear, and twist-like responses [Figs.~\ref{fig:zoo}(a), (b), and (d), respectively)]exhibit clear hinge domains with allosteric pathways typically lying along the boundary between these domains.
We also find a buildup of pure allosteric pathways [Fig.~\ref{fig:zoo}(d)] with no significant hinge domains at $\tau_h/\delta u^* \approx 0$ with $\tau_p/\delta u^* > 0$.
Finally, we find instances of networks with what we call ``isostatic responses'' [Fig.~\ref{fig:zoo}(e)], in which neither a significant allosteric pathway nor a pair of hinge domains is observed.
In these networks, the input response decays rapidly into the bulk of the network, before being amplified at the target, with very little stress communicated through the network~\cite{Yan2017}.
We note that the examples shown in Fig.~\ref{fig:zoo} are chosen to correspond to archetypal responses identified in the protein allostery literature. We stress that they are \emph{atypical}; most networks cannot be characterized by a single archetype but instead exhibit a combination of multiple archetypal motions. 

To assess the functional significance of the identified hinge domains and strain pathways, we measure the robustness of the tuned response to changes in local structure near these identified features.
Specifically, we measure the change in the strain of the target nodes from the tuned response $\Delta \varepsilon_T$ when a node and its nearest neighbors (those with which it shared an edge in the original untuned network structure) are rigidly constrained so that they may only collectively undergo pure rigid-body translations and rotations.
We then plot the average magnitude of the fractional change, $|\Delta \varepsilon_T /\varepsilon_T|$, as a function of the distance from the allosteric pathway or hinge domain boundaries.
We find that the average change in response is greatest for nodes that lie along an allosteric pathway (blue) or hinge boundary (red), implying that the nodes associated with these features tend to be most important for transmitting strain from the source to the target.

\section*{Analysis of Allosteric Proteins}

Now that we have developed and refined a topological analysis of allosteric response in tuned central-force spring networks, we apply it to a collection of allosteric proteins chosen from the Protein Data Bank (PDB)~\cite{Berman2000}. 
We consider proteins for which crystalline x-ray structures have been identified corresponding to both their inactive and active states (unbound and bound to substrate, respectively)~\cite{Daily2007, Yuan2015, Mitchell2016, Thirumalai2019}.
For each protein, we use the finite displacements of all atoms between the inactive and active configurations to compute the LRMSD.
Since our topological analysis requires a network structure, we utilize the Arpeggio Python package~\cite{Jubb2017} to construct an approximate atomic contact network for each protein.
We then apply our analysis to identify a pair of hinge domains and an allosteric pathway between the allosteric site (the source) where the effector ligand binds and the active site (the target) where the substrate binds (see Methods).
Software to 

Fig.~\ref{fig:proteins} shows the results of this analysis for some examples of allosteric proteins. 
For each protein, we depict the configurations before (top row) and after (bottom row) the binding of the effector ligand(s) at the allosteric site(s), which we treat as the undeformed and deformed configurations, respectively.
In the top row, we show the finite atomic displacements exhibited by each protein,
while in the bottom row, we show the LRMSD, along with the most topologically significant hinge domains and allosteric pathway identified by our analysis.

In Fig.~\ref{fig:proteins}(a), we see that when glucokinase is bound by a synthetic activator at the allosteric site, it undergoes a scissor-like hinge motion, causing it to bind a glucose molecule at the active site.
Our analysis successfully captures this hinge motion, dividing the protein into upper and lower hinge domains, while also identifying an allosteric pathway between the allosteric and active sites located along the boundary between the domains. 
Fig.~\ref{fig:proteins}(b) depicts the inhibition of aspartate carbamoyltransferase (ATCase) via CTP.
While to the eye, the motion of this protein does not display a clear hinge-like motion, 
our analysis does identify a pair of domains, separated along a plane of local deformations, with the allosteric pathway lying along this boundary.
The identification of this plane of local deformations is consistent with previous studies~\cite{Mitchell2016}.

\begin{figure}[t!]
\centering
\includegraphics[width=\linewidth]{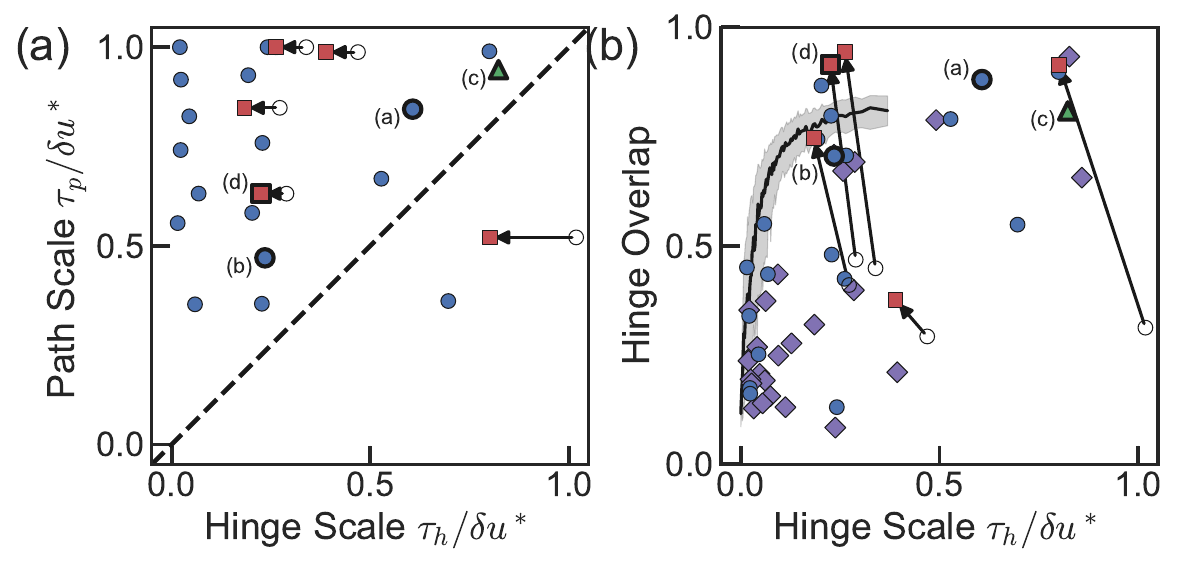}
\caption{
(a) Hinge and path scales for allosteric proteins. 
(b) Overlap of allosteric motion with approximate motion of rigid hinge domains as a function of the hinge scale for allosteric proteins (points) and tuned mechanical networks (black curve).
Error bands depict interquartile range at each value of the hinge scale.
In both panels, proteins containing an allosteric pathway between the allosteric and active sites are shown as blue circles, proteins containing a cooperative pathway are shown as green triangles, and proteins for which neither could be identified are shown with magenta diamonds.
Proteins for which more than two hinge domains were identified are shown with red squares with arrows indicating the difference in values compared to decomposing the protein into only two hinge domains.
The four labeled points in each panel correspond to the proteins in Fig.~\ref{fig:proteins}.
}
\label{fig:stats}
\end{figure}

In Fig.~\ref{fig:stats}(a), we plot the hinge and path scales for each protein in conformational phase space.
The points labeled (a)-(d) correspond to the proteins depicted in Fig.~\ref{fig:proteins}.
Much like our computer-generated allosteric networks, we find that proteins tend to occupy the upper left diagonal region, indicating that a large strain pathway scale tends to accompany a large hinge scale. Moreover, proteins with strong scissor-like, hinge or shear deformations tend to lie near the diagonal at relatively high hinge and pathway scales, similar to tuned central-force spring networks that display those behaviors.
For instance, glucokinase [the point labeled (a)] demonstrates relatively large hinge and pathway scales somewhat above the diagonal, similar to the central-force spring network that exhibited scissor-like deformation in Fig.~\ref{fig:zoo}(a).
Also, ATCase [the point labeled (b)] does not have a particularly strong hinge scale; this is why it is not easy to pick out hinge-like motion by eye. 

In Fig.~\ref{fig:stats}(b), we also measure the overlap of the allosteric protein motions with the approximation to the response obtained by treating each hinge domain as a completely rigid body (see Methods).
We plot this overlap as a function of the hinge scale and compare to our synthetic systems.
We find that in both cases, as the hinge scale increases, the overall average total overlap rapidly increases,
indicating that the hinge domains we identify account for most of the motion even for moderate values of the hinge scale. This result implies that for sufficiently high hinge scales, the features identified by our analysis fully characterize the allosteric response at a quantitative level.

Finally, we apply two different extensions of our analysis. 
First, we identify pathways between different active sites, 
capturing the cooperative effects between the binding of multiple substrates.
Fig.~\ref{fig:proteins}(c) depicts the classic example of hemoglobin in which the binding of oxygen at one heme group facilitates binding at the three other heme groups.
In this case, we show a cooperative pathway identified between a pair of oxygen binding sites.
In Fig.~\ref{fig:stats}(a), when no allosteric site is identified in a configuration, but multiple substrate binding sites are present like in hemoglobin, we plot the path scale for the most topologically significant cooperative pathway between all pairs of substrate binding sites.

Second, our analysis can be used to separate proteins into more than two hinge domains.
For example, in Fig.~\ref{fig:proteins}(d), we find that D-3-phosphoglycerate dehydrogenase (PGDH) is well-described by four hinge domains that do not correspond to the four monomer subunits.
For such multidomain hinges, we measure topological significance via the smallest hinge scale observed between all pairs of adjacent domains (see Methods).
In Fig.~\ref{fig:stats}, we show the change in values for proteins where allowing for additional hinge domains increases the hinge overlap significantly with a minimal decrease in the hinge scale.

\section*{Discussion}

We have used topological data analysis to analyze the responses of tuned allosteric mechanical networks in terms of hinge domains and allosteric pathways. We find that our tuned networks do not fall into discrete clusters corresponding to specific archetypal motions. Instead, they
exhibit allosteric conformational changes that exist along a continuum of hinge and pathway scales. With this insight in hand, we then applied our analysis to a collection of proteins, showing that our findings extend to proteins, opening up several avenues for future study.

The method we presented allows for the identification of unique hinge domains and strain pathways while providing a natural measure of each feature's topological significance without the need for choosing arbitrary cutoffs in the response.
Furthermore, it can be used to identify cooperative pathways between different active sites, and more generally, between any arbitrary pair of atoms.
It can separate proteins into any desired number of hinge domains, not just two, and can assess the significance of each additional domain.

A minimal implementation of our analysis only requires node or atomic positions for the inactive and active allosteric conformations.
In this study, we chose to use the underlying interactions via the spring network of each \emph{in silico} system or atomic contact networks of each protein.
However, our method can also be implemented without the need for such information.
For example, one could generate a network by simply adding edges between nodes or atoms below some cutoff distance or by constructing a Delaunay triangulation.

We chose to focus primarily on the allosteric pathways and hinge domains with the largest path and hinge scales, respectively. 
However, our analysis more generally yields a unique pathway between every pair of nodes, along with a large number of candidate hinge domains of varying significance.
One could imagine using different selection criteria to find features with different properties. For example, one could choose hinge domains with maximum hinge overlap.
Alternatively, a trained eye might use the features to guide intuition.

In our tuned networks we found that nodes close to allosteric pathways or hinge boundaries tend to be the most important for communicating allosteric responses.
In the context of proteins, this result suggests that the amino acids in such regions may be prime targets for controlling the behavior of allosteric proteins, or as potential drug targets~\cite{Nussinov2013, Dokholyan2016, Guarnera2016}. 

It would be interesting to understand whether the features identified by our analysis correspond to those that emerge from analyses based  on sequence data~\cite{Halabi2009} or low energy modes~\cite{Wodak2019, Ravasio2019},
both of which have been shown to be mutually related~\cite{Wang2019}.
For example, the hinge domain boundaries or allosteric pathways may correspond to highly conserved amino acids which are responsible for encoding the allosteric deformation as a low-energy mode.

Finally, deep learning has proven to be extremely effective at determining folded protein structures from sequence data~\cite{Jumper2021}.
Using similar techniques to predict the motions of allosteric proteins, or potential allosteric and active sites, poses an interesting challenge.
Since our topological analysis provides an informative set of features that capture allosteric function,
it may prove useful for constructing training sets for such machine learning-based approaches. 
Such methods also prove useful for proteins for which x-ray crystal structures are typically difficult to obtain, such as trans-membrane proteins, which often rely on allosteric signaling~\cite{Kermani2021}.

The most important consequence of our analysis, however, is that it provides a unifying framework for quantitatively characterizing allosteric response. The myriad and often complex allosteric responses found in tens of thousands of tuned central-force spring networks, as well as those of a score of proteins, can all be characterized in the same way, in terms of just two numbers -- the hinge and pathway scales.  Our results therefore provide a new view of allostery in proteins, freeing us from categorizing mechanisms in terms of distinct archetypes and directing us instead towards hinge domains and pathways extracted by threshold-free topological data analysis.

\matmethods{

\subsection{Creation of Tuned Allosteric Networks}

\subsubsection{Linear Response of Mechanical Networks} 
Here, we use the same linear response formulation as in Ref.~\cite{Rocks2019} for central-force spring networks with $N$ nodes and $N_b$ pairwise bonds.
The response of each node $i$ to external forces is represented by a $d$-dimensional displacement vector $\vec{u}_i$.
The strength of each bond $\langle i, j \rangle$ linking nodes $i$ and $j$ is characterized by a stiffness $k_{ij} = \lambda_{ij}/\ell_{ij}$, where $\lambda_{ij}$ is the stretch modulus per unit length and $\ell_{ij}$ is the rest length.

To calculate the elastic response, we minimize the deformation energy, which can be written in the absence of prestresses to quadratic order in the displacements as
\begin{equation}
E = \frac{1}{2}\sum_{\langle i, j\rangle}k_{ij}\qty[\hat{b}_{ij}\cdot \Delta \vec{u}_{ij}]^2,\label{eq:energy}
\end{equation}
where $\Delta \vec{u}_{ij} = \vec{u}_j - \vec{u}_i$ and  $\hat{b}_{ij}$ is a unit vector pointing from node $i$ to node $j$.
In matrix form, \eqref{eq:energy} is written as
\begin{equation}
E = \frac{1}{2}\mel{u}{\mathbf{H}}{u}
\end{equation}
where $\ket{u}$ is a length-$dN$ vector containing the displacements of all the nodes and $\mathbf{H}$ is the size-$dN\times dN$ Hessian matrix.
The Hessian can be expressed as
\begin{equation}
\mathbf{H} = \mathbf{Q}\mathbf{K}\mathbf{Q}^T
\end{equation}
where the compatibility matrix $\mathbf{Q}$ of size $dN\times N_b$ maps deformations of nodes to bonds and $\mathbf{K}$ is a size-$N_b\times N_b$ diagonal matrix of stiffnesses for each bond.

In this work, we are interested in the network response when fixed strains are imposed on $N_S$ pairs of source nodes. 
We define the strain on a pair of nodes $i$ and $j$ as
\begin{equation}
\varepsilon_{S, ij} =  \hat{b}_{ij}\cdot \Delta \vec{u}_{ij}/\ell_{ij}\label{eq:strain}
\end{equation}
in the linear limit and organize them in a length-$N_S$ vector $\ket{\varepsilon_S}$.
We note that this definition does not require the nodes to share a bond.
We define the $dN\times N_S$ matrix $\mathbf{S}$ to measure the strains on the source nodes according to \eqref{eq:strain}, $\ket{\varepsilon_S} = \mathbf{S}^T\ket{u}$.
We then calculate the response using the Lagrangian
\begin{equation}
\mathcal{L} = E - \bra{\sigma_S}(\ket{\varepsilon_S} - \ket{\varepsilon_S^*}) -  \mel{\lambda_G}{\mathbf{G}^T}{u}\label{eq:lagrangian}
\end{equation}
where $\ket*{\varepsilon_S^*}$ is a length-$N_S$ vector of the desired imposed strains and $\ket{\sigma_S}$ is a corresponding vector of Lagrange multipliers for each constraint.
Since the Lagrange multipliers appear conjugate to $\ket{\varepsilon_S}$, we interpret $\ket{\sigma_S}$ as the necessary stresses needed to obtain the desired strains.
We also add an additional set of $d(d+1)/2$ constraints with Lagrange multipliers $\ket{\lambda_G}$ which constrain the global translational and rotational degrees of freedoms to zero, represented by a $dN\times d(d+1)/2$ matrix $\mathbf{G}$.
Extremizing \eqref{eq:lagrangian}, we find the solution
\begin{equation}
\mqty(\ket{u}\\\ket{\sigma_S}\\\ket{\lambda_G}) = \mqty(\mathbf{H} & -\mathbf{S}^T & -\mathbf{G}^T\\
-\mathbf{S} & 0 & 0\\
-\mathbf{G} & 0 & 0)^{-1}\mqty(0\\-\ket{\varepsilon_S^*}\\0).
\end{equation}

\subsubsection{Generation of Initial Disordered Spring Networks}

Our initial spring networks are derived from jammed packings of soft spheres, created using standard jamming algorithms: particles are initially placed at random in a $d$-dimensional periodic box, quenched to a local energy minimum, and then adjusted until the pressure is approximately $p=10^{-1}$~\cite{OHern2003, Liu2010, Goodrich2016}.
In this study, we consider packings in $d=3$ with harmonic interactions, consisting of a 50:50 bi-disperse mixture of particles with diameter ratio $1.0\sigma:1.4\sigma$ where $\sigma$ is the diameter of the smaller particles.
We generated a total of $4096$ independent systems each consisting of $2048$ particles.

To convert a jammed packing into a central-force spring network, we replace each particle with a node and each particle contact with an unstretched spring of unit stretch modulus $\lambda_{ij}=1$.
We then create a ball-shaped network by removing nodes (and adjacent edges) beyond a cutoff radius $R_{\mathrm{cutoff}}$ chosen to result in a network containing approximately 512 nodes, ${\frac{4}{3}\pi R_{\mathrm{cutoff}}^3/L^3 = 512/2048}$ (where $L^3$ is the box volume).
To ensure the resulting network is rigid, we eliminate any zero modes by repeatedly calculating the lowest energy mode of the Hessian and removing the node whose motion contributes most to this mode, 
stopping when the lowest eigenvalue of the Hessian is at least $10^{-2}$.
Finally, to ensure no zero modes will be introduced during the network tuning process, 
we introduce additional 2nd nearest neighbor interactions of stiffness $k_{ij} = 10^{-4}$ that we hold fixed while tuning,
and are maintained in the final tuned network.
In the final ensemble, each network has on average 509$\pm$ 9 nodes with 2070$\pm$ 37 nearest neighbor bonds (see SI for distributions).

Next, we choose the source and target nodes from the collection of surface nodes,
which we define as those nodes that had at least one edge removed when the periodic network was converted to a ball-shaped one.
We pick a random point on the surface of the network and choose the first source node as the surface node closest to the point,
and the second source node as the next closest surface node that does not share an edge with the first.
We choose the target nodes by repeating this process at the position on the opposite pole of the network.
By choosing each pair of nodes so that they do not share an edge,
we increase the likelihood that forces applied to these nodes will couple to the rest of the system.

\subsubsection{Tuning Protocol for Allosteric Responses}

In this work, we consider allosteric responses in which the target nodes exhibit a strain  $\varepsilon_T \leq -1$ when a strain of size $\varepsilon_S = 1$ is applied to to the source nodes.
To create mechanical networks with such responses, we optimize the spring networks we generate using a greedy algorithm in which we add and remove bonds.

Starting with the untuned network, at each step, we iterate through each bond in the network (excluding 2nd nearest neighbors interactions) and determine how the response changes when that bond is either removed completely or added back in if it was removed in a previous step.
For each bond, we measure the resulting strain of the target nodes $\varepsilon_T$, along with the corresponding stress $\sigma_T$ that could potentially be exerted by those nodes.
To measure the target strain $\varepsilon_T$, we constrain the source nodes to the source strain $\varepsilon_S$ and measure the response of the target.
To measure the target stress $\sigma_T$, we apply the source strain while simultaneously constraining the target nodes to exhibit zero strain.
We then measure the target stress $\sigma_T$ via the negative of the Lagrange multiplier used to constrain the target bonds [see \eqref{eq:lagrangian}].
Next, we choose to remove or add the edge that results in a negative stress with the largest magnitude, by setting $\lambda_{ij}=0$ or $\lambda_{ij}=1$, respectively.
We repeatedly modify bonds using this process until either the target strain is $\varepsilon_T \leq -1$ (within a margin of $10^{-8}$) or changes by less than $10^{-4}$. 

We cycle through each of the 4096 spring networks in our ensemble (choosing different random source and target nodes each time) to achieve $~10^5$ different tuning attempts.
Of these, we find that only about $35\%$ of networks are successfully tuned to the desired target strain, resulting in about 35000 allosteric networks.
Unlike previous studies, our success rate is not around $100\%$ because we tune the strain indirectly via the stress. If we were to tune the strain directly, we would achieve a success rate close to $100\%$, but we find that most of that network would not exhibit significant force propagation through the network~\cite{Rocks2017}.

\subsection{Topological Data Analysis}

\subsubsection{Local Root-Mean-Square Displacement (LRMSD)}

We characterize the local deformations associated with allosteric responses via the local root-mean-square displacement (LRMSD),
which measures the deviation of node displacements in a local neighborhood from rigid-body translations and rotations.
To calculate the LRMSD $\delta u_i$ around node $i$, we first find the local $d\times d$ deformation tensor $\hat{\mathbf{F}}_i$ that best approximates the motion,
\begin{equation}
\hat{\mathbf{F}}_i = \argmin_{\mathbf{F}}\sum_j w_{ij}\qty[\vec{b}_{ij} + \Delta \vec{u}_{ij} - \mathbf{F}\vec{b}_{ij} ]^2\label{eq:defmat},
\end{equation}
where $\vec{b}_{ij}$ is a vector pointing from node $i$ to $j$ in the undeformed configuration.
To pick out the local neighborhood, we weigh each particle by its distance from node $i$ using Gaussian weights,
\begin{equation}
w_{ij} =  \frac{e^{-\frac{\ell_{ij}^2}{2\sigma^2}}\Theta(\ell_{\max}-\ell_{ij})(1-\delta_{ij})}{\sum\limits_k e^{-\frac{\ell_{ik}^2}{2\sigma^2}}\Theta(\ell_{\max}-\ell_{ik})(1-\delta_{ik})},
\end{equation}
where $\ell_{ij}$ is the distance between $i$ and $j$ in the undeformed configuration, $\sigma$ is a characteristic length scale, and $\ell_{\max}$ is a cutoff distance which we choose to be $3\sigma$.
For our random spring networks, we choose $\sigma$ to be the diameter of the smaller particles in the original jammed packings.
The resulting deformation tensor around particle $i$ is then
\begin{equation}
\hat{\mathbf{F}}_i = \qty[\sum_j w_{ij}\qty(\vec{b}_{ij} + \Delta \vec{u}_{ij})\vec{b}_{ij}^T]\qty[\sum_k w_{ik} \vec{b}_{ik}\vec{b}_{ik}^T]^{-1}.
\end{equation}
We then use $\hat{\mathbf{F}}_i$ to estimate the local rigid-body rotation matrix $\mathbf{R}_i$.
For nonlinear deformations, we use the formula ${\mathbf{R}_i = \hat{\mathbf{F}}_i (\hat{\mathbf{F}}_i^T\hat{\mathbf{F}}_i)^{-\frac{1}{2}}}$,
while for linear deformations, we use ${\mathbf{R}_i \approx \mathbf{I} +   \frac{1}{2}(\hat{\mathbf{F}}_i-\hat{\mathbf{F}}_i^T)}$. 
Finally, we define the LRMSD as
\begin{equation}
\delta u_i = \sqrt{\sum_j w_{ij}\qty[\Delta \vec{u}_{ij} - (\mathbf{R}_i-\mathbf{I})\vec{b}_{ij} ]^2}.
\end{equation}

\subsubsection{Identification of Hinge Domains and and Strain Pathways}

To analyze the topological structure of allosteric responses, we utilize the methods for topological coarse-graining developed in Ref.~\cite{Rocks2020}.
In this approach, the response $\delta u_i$ defined on the nodes of a network is used to create a pair of spanning trees,
which we call the hinge and pathway skeletons.
These two trees are then used to identify hinge domains or allosteric pathways, respectively.
For our mechanical networks, we utilize the network of all nearest-neighbor interactions present before tuning, even if they are eventually removed by the tuning process.
We provide examples of both the hinge and pathway skeletons of a network in the SI.

To create the hinge skeleton, we first use $\delta u_i$ on the nodes to introduce an ordering on the edges of the network.
For each edge connecting nodes $i$ and $j$, we create a tuple $(\delta u_i, \delta u_j)$, sorted so that $\delta u_i > \delta u_j$,
and then sort these tuples in ascending lexicographic order.
We then use this ordering to perform a filtration of $\delta u_i$ on the network.
Starting with an empty network with no edges, we add each edge to the network in order, one at a time,
taking note of any changes in the connected components, defined as contiguously connected sets of edges.
In particular, we focus on changes in the number of connected components or their sizes (defined as the number of nodes contained in a component). 

When a new edge is introduced, there are four possibilities: 
(i) the new edge is not connected to any of the pre-existing edges, creating a new connected component, 
(ii) the new edge is shared between two different pre-existing components and joins them together,
(iii) the new edge is connected to a single pre-existing component by a single node, incurring no change in the number of connected components, but increases its size,
or (iv) the new edge is connected to a single pre-existing component by both nodes.
We define the hinge skeleton as the collection of edges defined by (i), (ii), and (iii).
Analogously, to create the pathway skeleton, we perform the same process with a reverse ordering of the nodes, equivalent to replacing $\delta u_i$ with $-\delta u_i$ throughout the process.

To identify a pair of hinge domains from the hinge skeleton,
we utilize the edges that fall into case (ii) above, which we call boundary edges.
These boundary edges uniquely divide the hinge skeleton into a set of domains, each containing a single local minimum in $\delta u_i$.
Each domain corresponds to a region of the network that moves with relatively rigid motion compared to its surroundings.
Since the hinge skeleton is a tree, removing a single boundary edge from the network creates a pair of domains.
To assess the topological significance of such domains, we first measure the maximum LRMSD of the nodes that comprise the boundary edge, $\delta u_{\text{boundary}}$.
This quantity characterizes the minimal maximum strain of any path in the full network that goes from one of the domains to the other, i.e., the minimal deformation separating the two domains along the boundary.
We then measure the minimum LRMSD in each of the two domains and take the maximum of these two values, $\delta u_{\text{domain}}$.
We define the hinge scale as the difference between these two values,
\begin{equation}
\tau_h = \delta u_{\text{boundary}} - \delta u_{\text{domain}}.
\end{equation}
This quantity is an approximation of -- and often equivalent to -- the persistence of the connected component corresponding to $\delta u_{\text{domain}}$.
We measure the topological significance via $\tau_h$ for each candidate pair of domains created by a boundary edge, choosing the domains with the largest value.
We skip any of the boundary edges that would create a sector smaller than a minimal cutoff $N_{\min}$.
For mechanical networks, we choose $N_{\min} = 3$, the minimal number of nodes needed to uniquely calculate the rigid body motion of a domain in three dimensions.

To identify an allosteric pathway of maximal strain between a source node and a target node, we simply identify the unique pathway on the pathway skeleton between those nodes.
To assess the topological significance of a pathway, we measure the minimum LRMSD encountered along the path,  $\delta u_{\text{min}}$.
We then define the strain path scale as
\begin{equation}
\tau_p = \delta u_{\text{min}}.
\end{equation}
In our analysis, we choose the pathway with the largest value of $\tau_p$.
If two pathways share the same value, we choose the shorter of the two.
For mechanical networks, we choose the normalization scale  $\delta u^*$ as the LRMSD of the source node at the beginning of the chosen allosteric pathway.

\subsubsection{Hinge Motion}

To characterize the motion of the hinge domains we identify,
we assess the similarity of their motion to that of two rigidly moving hinge domains.
To find this approximate rigid-body response for a single hinge domain,
we first calculate the position and displacement of the center of mass,
\begin{align}
\vec{X}_{cm} &= \frac{1}{N_d}\sum_i \vec{X}_i, &
\vec{u}_{cm} &= \frac{1}{N_d}\sum_i \vec{u}_i
\end{align}
where $\vec{X}_i$ is the position of node $i$ in the undeformed coordinates and the sum ranges over the nodes in the hinge domain of size $N_d$.

Next, we calculate the approximate rigid-body rotation experienced by the hinge domain.
Similar to our calculation of the LRMSD, we find the deformation matrix $\hat{\mathbf{F}}$ via  \eqref{eq:defmat} which best approximates the overall motion.
However, we perform this calculation relative to the center of mass, rather than particle $i$, and take the weights to be unity for all particles with no cutoff distance.
We then extract the rotation matrix $\mathbf{R}$ from $\hat{\mathbf{F}}$ in the same way.
The approximate rigid-body motion of each particle in the hinge domain is then
\begin{equation}
\vec{u}_{h, i} = \vec{u}_{cm} +  (\mathbf{R} - \mathbf{I})(\vec{X}_i-\vec{X}_{cm}).
\end{equation}
We assemble these displacements for all particles in all hinge domains into a single length-$dN$ vector $\ket{u_h}$ and then measure the similarity of this motion to the original displacement $\ket{u}$ via the hinge overlap,
\begin{equation}
q = \frac{\braket{u}{u_h}}{\norm*{u}\norm*{u_h}}.
\end{equation}

\subsection{Analysis of Allosteric Proteins}

We assembled a database of 55 allosteric protein structures from Refs.~\cite{Thirumalai2019}, \cite{Daily2007}, \cite{Mitchell2016}, and \cite{Yuan2015}.
For each protein, we obtained a pair of active (bound to substrate) and inactive (not bound to substrate) configurations from the Protein Data Bank (PDB)~\cite{Berman2000},
and manually identified the relevant ligands and substrates corresponding to the allosteric and active sites, respectively.
For proteins that undergo phosphorylation at the allosteric site, we treat the phosphorylated residues as effective ligands.
We also ignore any atoms with insertion codes specifying alternative residues. An annotated table of these proteins is provided in the SI.

We construct an approximate atomic contact network for each fully assembled protein structure and any relevant ligands or substrates using the Arpeggio Python package~\cite{Jubb2017} with default settings.
We utilize all contact types identified by the software excluding proximal contacts.
We identify allosteric and active sites on a protein structure by recording any atoms in contact with a ligand or substrate molecule.
Before we perform any analysis, we remove any atoms or contacts associated with any ligands or substrates since they will not be shared between the active and inactive configurations.

To calculate displacements, we identify corresponding atoms in inactive and active protein structures, excluding any bound ligands or substrates.
We then calculate the difference in positions using the active configuration as the reference.
This means that if a  protein is activated by the binding of ligand(s) to the allosteric site(s), then we utilize the configuration with bound ligand as the reference configuration.
Alternatively, if the protein is inhibited, we utilize the configuration without bound ligand as the reference.
We subtract any global rigid-body translations and rotations assuming nonlinear deformations (see Hinge Motion). 
Next, we calculate the LRMSD for each atom in the protein using a characteristic length scale of $\sigma = 5\AA$.

When identifying hinge domains, we use a minimum domain size of $N_{\min} = 200$ atoms, or approximately 10 amino acids.
This is to ensure that small chains of atoms on the protein surface that exhibit large deformations are not identified as hinge domains.
To divide proteins into more than two hinge domains, we iteratively repeat our hinge identification procedure, removing the boundary edges chosen by the procedure from the hinge skeleton at each step.

To find allosteric pathways, we consider every possible pair of atoms between each allosteric and active site.
For cooperative strain pathways, we perform the same analysis between each pair of atoms between active sites corresponding to different substrate molecules.

To normalize the hinge and path scales, we choose a normalization scale $\delta u^*$ depending on which information is available in the protein structures.
If we are able to identify both an allosteric and active site, we choose $\delta u^*$ to be the LRMSD of the atom at the beginning of the pathway at the allosteric site.
If a protein only has active sites labeled, we instead choose $\delta u^*$ as the maximum LRMSD of the atoms at the ends of the strain pathway at the active site.
If a protein does not have any labeled active sites, we choose $\delta u^*$ as the maximum LRMSD of all atoms at the allosteric site(s).
Alternatively, if a protein does not have any labeled allosteric sites, we choose $\delta u^*$ as the maximum LRMSD of all atoms at the active site(s).

\subsection{Code and Data Availability}
We provide a GitHub repository containing code to perform our protein analysis at \href{https://github.com/jrocks/allostery_tda}{https://github.com/jrocks/allostery\_tda}.
Files containing the list of proteins analyzed in this paper and their analysis results, 
along with a database of tuned mechnical networks is also provided at (TBD).
Descriptions of the data files are provided in the SI.
}

\showmatmethods{} 

\acknow{
We thank Stanislas Leibler for instructive discussions that initiated our studies of allostery, and Sid Nagel for an inspiring collaboration on tuning and training allostery in mechanical networks that made this work possible.
This work was supported by the 
NIH NIGMS through 1R35GM119461 (J.W.R.), the NSF through DMR-2005749 (A.J.L.) and the UPenn MRSEC DMR-2309043 (E.K.),
and the Simons Foundation through Investigator grants \#327939 (J.W.R. and A.J.L.) and \#568888 (E.K.).
The authors also acknowledge support from the Shared
Computing Cluster administered by Boston University Research Computing Services.
A.J.L. thanks CCB at the Flatiron Institute (a division of the Simons Foundation), as well as the Isaac Newton Institute for Mathematical Sciences under the program ``New Statistical Physics in Living Matter" (EPSRC grant EP/R014601/1), for support and hospitality while a portion of this research was carried out.
}

\showacknow{} 

\bibliography{allostery}

\begin{thebibliography}{10}

\bibitem{Ribeiro2016}
AAST Ribeiro, V Ortiz, {A Chemical Perspective on Allostery}.
\newblock {\em\protect\JournalTitle{Chemical Reviews}} \textbf{116}, 6488--6502
  (2016).

\bibitem{Thirumalai2019}
D Thirumalai, C Hyeon, PI Zhuravlev, GH Lorimer, {Symmetry, Rigidity, and
  Allosteric Signaling: From Monomeric Proteins to Molecular Machines}.
\newblock {\em\protect\JournalTitle{Chemical Reviews}} \textbf{119}, 6788--6821
  (2019).

\bibitem{Wodak2019}
SJ Wodak, et~al., {Allostery in Its Many Disguises: From Theory to
  Applications}.
\newblock {\em\protect\JournalTitle{Structure}} \textbf{27}, 566--578 (2019).

\bibitem{Daily2007}
MD Daily, JJ Gray, {Local motions in a benchmark of allosteric proteins}.
\newblock {\em\protect\JournalTitle{Proteins: Structure, Function and
  Genetics}} \textbf{67}, 385--399 (2007).

\bibitem{Shatsky2004}
M Shatsky, R Nussinov, HJ Wolfson, {FlexProt: Alignment of Flexible Protein
  Structures Without a Predefinition of Hinge Regions}.
\newblock {\em\protect\JournalTitle{Journal of Computational Biology}}
  \textbf{11}, 83--106 (2004).

\bibitem{Keating2009}
KS Keating, SC Flores, MB Gerstein, LA Kuhn, {StoneHinge : Hinge prediction by
  network analysis of individual protein structures}.
\newblock {\em\protect\JournalTitle{Protein Science}} \textbf{18}, 359--371
  (2009).

\bibitem{Taylor2014}
D Taylor, G Cawley, S Hayward, {Quantitative method for the assignment of hinge
  and shear mechanism in protein domain movements}.
\newblock {\em\protect\JournalTitle{Bioinformatics}} \textbf{30}, 3189--3196
  (2014).

\bibitem{Mitchell2016}
MR Mitchell, T Tlusty, S Leibler, {Strain analysis of protein structures and
  low dimensionality of mechanical allosteric couplings.}
\newblock {\em\protect\JournalTitle{Proceedings of the National Academy of
  Sciences of the United States of America}} \textbf{113}, E5847--E5855 (2016).

\bibitem{Feher2014}
VA Feher, JD Durrant, AT {Van Wart}, RE Amaro, {Computational approaches to
  mapping allosteric pathways} (2014).

\bibitem{Dokholyan2016}
NV Dokholyan, {Controlling Allosteric Networks in Proteins}.
\newblock {\em\protect\JournalTitle{Chemical Reviews}} \textbf{116}, 6463--6487
  (2016).

\bibitem{Rocks2020}
JW Rocks, AJ Liu, E Katifori, {Revealing structure-function relationships in
  functional flow networks via persistent homology}.
\newblock {\em\protect\JournalTitle{Physical Review Research}} \textbf{2},
  033234 (2020).

\bibitem{Rocks2021}
JW Rocks, AJ Liu, E Katifori, {Hidden Topological Structure of Flow Network
  Functionality}.
\newblock {\em\protect\JournalTitle{Physical Review Letters}} \textbf{126},
  028102 (2021).

\bibitem{Flechsig2017}
H Flechsig, {Design of Elastic Networks with Evolutionary Optimized Long-Range
  Communication as Mechanical Models of Allosteric Proteins}.
\newblock {\em\protect\JournalTitle{Biophysical Journal}} \textbf{113},
  558--571 (2017).

\bibitem{Rocks2017}
JW Rocks, et~al., {Designing allostery-inspired response in mechanical
  networks}.
\newblock {\em\protect\JournalTitle{Proceedings of the National Academy of
  Sciences}} \textbf{114}, 2520--2525 (2017).

\bibitem{Tlusty2017}
T Tlusty, A Libchaber, JP Eckmann, {Physical Model of the Genotype-to-Phenotype
  Map of Proteins}.
\newblock {\em\protect\JournalTitle{Physical Review X}} \textbf{7}, 021037
  (2017).

\bibitem{Yan2017}
L Yan, R Ravasio, C Brito, M Wyart, {Architecture and coevolution of allosteric
  materials}.
\newblock {\em\protect\JournalTitle{Proceedings of the National Academy of
  Sciences}} \textbf{114}, 2526--2531 (2017).

\bibitem{Yan2018}
L Yan, R Ravasio, C Brito, M Wyart, {Principles for Optimal Cooperativity in
  Allosteric Materials}.
\newblock {\em\protect\JournalTitle{Biophysical Journal}} \textbf{114},
  2787--2798 (2018).

\bibitem{Rocks2019}
JW Rocks, H Ronellenfitsch, AJ Liu, SR Nagel, E Katifori, {Limits of
  multifunctionality in tunable networks}.
\newblock {\em\protect\JournalTitle{Proceedings of the National Academy of
  Sciences}} \textbf{116}, 2506--2511 (2019).

\bibitem{Berman2000}
HM Berman, et~al., {The Protein Data Bank}.
\newblock {\em\protect\JournalTitle{Nucleic Acids Research}} \textbf{28},
  235--242 (2000).

\bibitem{OHern2003}
CS O'Hern, LE Silbert, AJ Liu, SR Nagel, {Jamming at zero temperature and zero
  applied stress: The epitome of disorder}.
\newblock {\em\protect\JournalTitle{Physical Review E}} \textbf{68}, 011306
  (2003).

\bibitem{Liu2010}
AJ Liu, SR Nagel, {The Jamming Transition and the Marginally Jammed Solid}.
\newblock {\em\protect\JournalTitle{Annual Review of Condensed Matter Physics}}
  \textbf{1}, 347--369 (2010).

\bibitem{Goodrich2016}
CP Goodrich, AJ Liu, JP Sethna, {Scaling ansatz for the jamming transition}.
\newblock {\em\protect\JournalTitle{Proceedings of the National Academy of
  Sciences of the United States of America}} \textbf{113}, 9745--9750 (2016).

\bibitem{Edelsbrunner2010}
H Edelsbrunner, JL Harer, {\em {Computational topology : an introduction}}.
\newblock (American Mathematical Society), p. 241 (2010).

\bibitem{Otter2017}
N Otter, MA Porter, U Tillmann, P Grindrod, HA Harrington, {A roadmap for the
  computation of persistent homology}.
\newblock {\em\protect\JournalTitle{EPJ Data Science}} \textbf{6}, 17 (2017).

\bibitem{Yuan2015}
Y Yuan, MF Tam, V Simplaceanu, C Ho, {New Look at Hemoglobin Allostery}.
\newblock {\em\protect\JournalTitle{Chemical Reviews}} \textbf{115}, 1702--1724
  (2015).

\bibitem{Jubb2017}
HC Jubb, et~al., {Arpeggio: A Web Server for Calculating and Visualising
  Interatomic Interactions in Protein Structures}.
\newblock {\em\protect\JournalTitle{Journal of Molecular Biology}}
  \textbf{429}, 365--371 (2017).

\bibitem{Nussinov2013}
R Nussinov, CJ Tsai, {Allostery in Disease and in Drug Discovery}.
\newblock {\em\protect\JournalTitle{Cell}} \textbf{153}, 293--305 (2013).

\bibitem{Guarnera2016}
E Guarnera, IN Berezovsky, {Allosteric sites: remote control in regulation of
  protein activity}.
\newblock {\em\protect\JournalTitle{Current Opinion in Structural Biology}}
  \textbf{37}, 1--8 (2016).

\bibitem{Halabi2009}
N Halabi, O Rivoire, S Leibler, R Ranganathan, {Protein Sectors: Evolutionary
  Units of Three-Dimensional Structure}.
\newblock {\em\protect\JournalTitle{Cell}} \textbf{138}, 774--786 (2009).

\bibitem{Ravasio2019}
R Ravasio, et~al., {Mechanics of Allostery: Contrasting the Induced Fit and
  Population Shift Scenarios}.
\newblock {\em\protect\JournalTitle{Biophysical Journal}} \textbf{117},
  1954--1962 (2019).

\bibitem{Wang2019}
SW Wang, AF Bitbol, NS Wingreen, {Revealing evolutionary constraints on
  proteins through sequence analysis}.
\newblock {\em\protect\JournalTitle{PLOS Computational Biology}} \textbf{15},
  e1007010 (2019).

\bibitem{Jumper2021}
J Jumper, et~al., {Highly accurate protein structure prediction with
  AlphaFold}.
\newblock {\em\protect\JournalTitle{Nature}} \textbf{596}, 583--589 (2021).

\bibitem{Kermani2021}
AA Kermani, {A guide to membrane protein X‐ray crystallography}.
\newblock {\em\protect\JournalTitle{The FEBS Journal}} \textbf{288}, 5788--5804
  (2021).

\end{thebibliography}

\end{document}



\maketitle






\FloatBarrier


\begin{figure}
\centering
\includegraphics[width=0.5\linewidth]{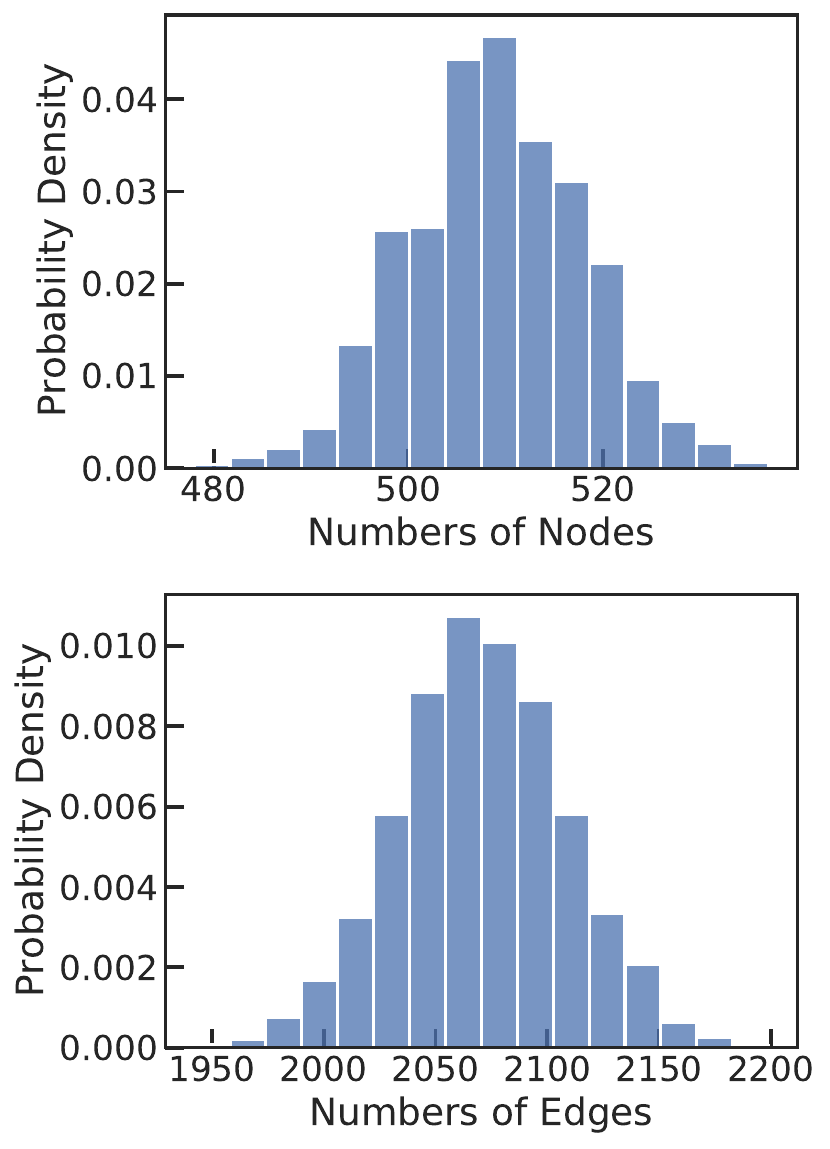}
\caption{Statistical properties of untuned network ensemble consisting of 4096 three-dimensional ball-shaped networks. Shown are the distribution of (a) nodes and (b) edges (nearest neighbor bonds). On average each network contains 590 $\pm$ 9 nodes with 2070 $\pm$ 37 edges. Error bars are measured using standard deviation.
}
\label{fig:network_dist}
\end{figure}

\FloatBarrier

\begin{figure}
\centering
\includegraphics[width=0.75\linewidth]{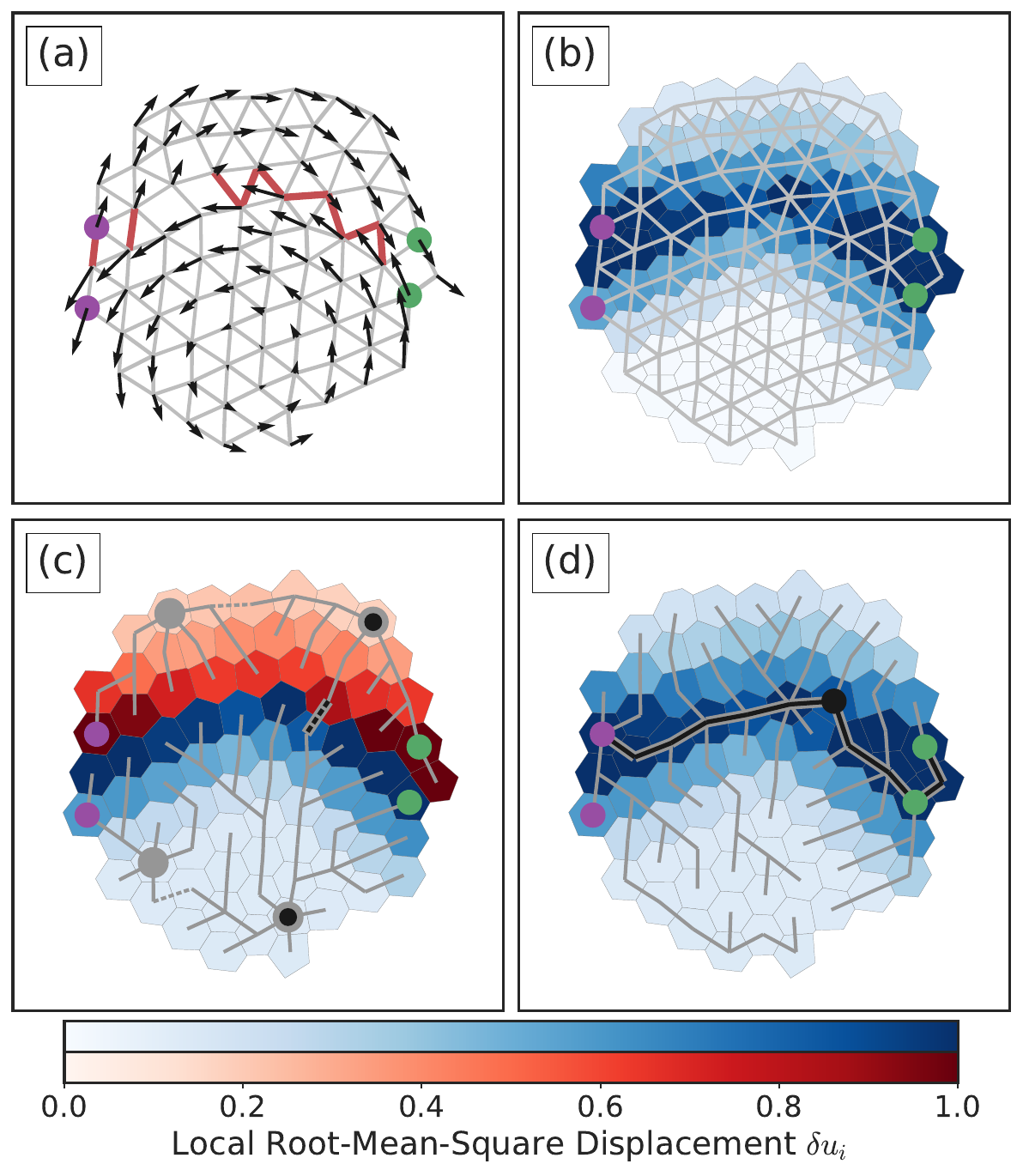}
\caption{Hinge and pathway skeletons. Examples are depicted for a two-dimensional mechanical network for visual clarity. 
The source and target nodes are shown in purple and green, respectively. (a) The tuned network structure is shown with removed bonds in red and the resulting displacements depicted with black arrows. 
(b) The local root-mean-square displacement $\delta u_i$ around each node $i$ is shown in shades of blue.
(c) The hinge skeleton is highlighted in grey, 
consisting of all edges comprising paths of steepest descent from each node to a local minimum in $\delta u_i$, indicated with a large grey circle. 
The hinge skeleton naturally decomposes the network into a collection of sectors, each corresponding to a different local minimum.
Boundaries between sectors are indicated by dashed edges.
The pair of hinge domains, shown in red and blue, are found by separating the hinge skeleton into two parts according to one of these edges, shown as a black dashed edge. 
The local minimum of each domain is shown as a black circle.
(d) The pathway skeleton is highlighted in grey, consisting of all edges comprising paths of steepest ascent from each node to a local maximum in $\delta u_i$. 
The path with the largest minimum strain between a source and target node is shown in black.
The node along the path with the minimum strain is highlighted with a black circle.
}
\label{fig:skeleton}
\end{figure}

\FloatBarrier

\dataset{network.shelve}{Database of tuned networks. Description and location TBD.}
\dataset{proteins.xlsx}{Table of proteins. Description and location TBD. Can currently be downloaded from:

 \href{https://github.com/jrocks/allostery_tda/blob/main/data/proteins.xlsx}{https://github.com/jrocks/allostery\_tda/blob/main/data/proteins.xlsx}.}
\dataset{results.xlsx}{Results of protein analysis. Description and location TBD. Can currently be downloaded from:

 \href{https://github.com/jrocks/allostery_tda/blob/main/data/results.xlsx}{https://github.com/jrocks/allostery\_tda/blob/main/data/results.xlsx}.}